\documentclass[aps,prl,showpacs,twocolumn,superscriptaddress,amsmath,amssymb]{revtex4}
\pdfoutput=1

\usepackage{graphicx}
\usepackage{dcolumn}
\usepackage{bm}

\preprint{APS/123-QED}

\begin{document}

\title{Experimental demonstration of single-site addressability\\ in a two-dimensional optical lattice}
\author{Peter W\"urtz}
\affiliation{Institut f\"ur Physik, Johannes Gutenberg-Universit\"at, 55099 Mainz, Germany}
\author{Tim Langen}
\affiliation{Institut f\"ur Physik, Johannes Gutenberg-Universit\"at, 55099 Mainz, Germany}
\author{Tatjana Gericke}
\affiliation{Institut f\"ur Physik, Johannes Gutenberg-Universit\"at, 55099 Mainz, Germany}
\author{Andreas Koglbauer}
\affiliation{Institut f\"ur Physik, Johannes Gutenberg-Universit\"at, 55099 Mainz, Germany}
\author{Herwig Ott}
\email{ott@uni-mainz.de}

\affiliation{Institut f\"ur Physik, Johannes Gutenberg-Universit\"at, 55099 Mainz, Germany}
\affiliation{Research Center OPTIMAS, Technische Universit\"at Kaiserslautern, 67663 Kaiserslautern, Germany}

\date{\today}

\begin{abstract}
We demonstrate single site addressability in a two-dimensional optical lattice with 600\,nm lattice spacing. After loading a Bose-Einstein condensate in the lattice potential we use a focused electron beam to remove atoms from selected sites. The patterned structure is subsequently imaged by means of scanning electron microscopy. This technique allows to create arbitrary patterns of mesoscopic atomic ensembles. We find that the patterns are remarkably stable against tunneling diffusion. Such micro-engineered quantum gases are a versatile resource for applications in quantum simulation, quantum optics and quantum information processing with neutral atoms. 
\end{abstract}


\maketitle
Ultracold atomic gases provide a versatile experimental platform to address fundamental quantum mechanical phenomena. On the one hand, they are a subject of research on their own \cite{Dalfovo1999,Ketterle2008}. On the other hand, they are a resource for applications in quantum optics \cite{Ginsburg2007}, quantum simulation \cite{Bloch2008} or quantum information processing \cite{Brennen1999,Raussendorf2001}. In order to use neutral atoms for such purposes, micro-structured quantum gases with suitable addressing schemes are essential. Optical tweezers \cite{Urban2008,Gaetan2008,Lengwenus2006}, optical conveyor belts \cite{Karski2009} and magnetic microtraps \cite{Fortagh2007,Whitlock2009} are currently investigated in this context. Ultracold quantum gases in optical lattices are an alternative as they can combine perfectly periodic potential structures in one, two and three dimensions with single site occupancy \cite{Greiner2002}. However, a lattice spacing below 1\,$\mu$m imposes stringent requirements on the spatial resolution of the addressing scheme. Several approaches are currently pursued in order to tackle this difficulty. Advanced optical addressing schemes are a natural choice \cite{Zhang2006,Gorshkov2008}. Coupling atoms to single ions \cite{Kollath2007} or a small solid probe are alternatives to reach this goal. Lattices with larger spacing \cite{Nelson2007} are also promising as they relax the requirements for the nominal resolution of the addressing scheme.

Here, we demonstrate single site addressability of a quantum gas in a 2D optical lattice with 600\,nm lattice spacing. To this end, we employ a recently developed imaging and manipulation technique based on scanning electron microscopy \cite{Gericke2008}. Loading a Bose-Einstein condensate in the lattice potential, we selectively remove atoms from individual sites by means of the dissipative interaction with a focused electron beam. In this way, arbitrary patterns of occupied lattice sites can be produced. We study the tunneling processes in the patterned structure and the depletion dynamics during the preparation stage.

\begin{figure}
	\centering
		\includegraphics[width=0.47\textwidth]{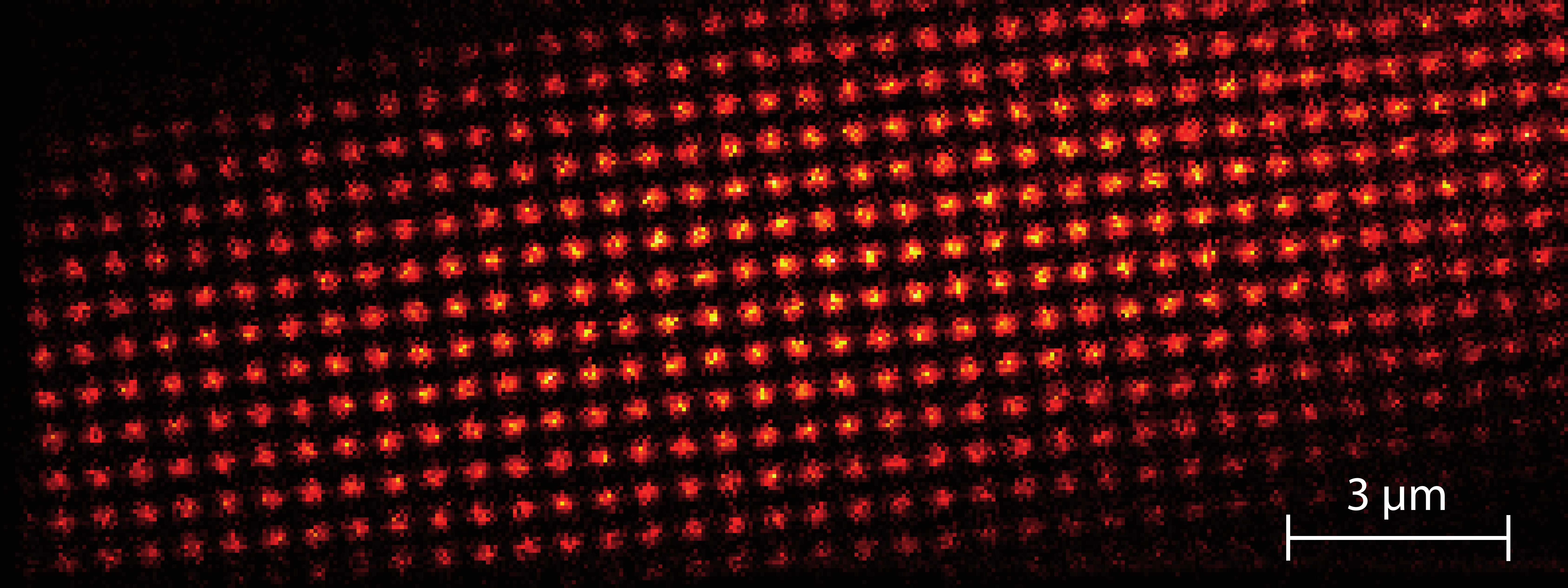}
	\caption{Electron microscope image of a Bose-Einstein condensate in a 2D optical lattice with 600\,nm lattice spacing (sum obtained from 260 individual experimental realizations). Each site has a tube-like shape with an extension of 6\,$\mu$m perpendicular to the plane of projection. The central lattice sites contain about 80 atoms.}
	\label{fig:figure1}
\end{figure}

In our experiment we use a scanning electron microscope which is implemented in a standard apparatus for the production of ultracold quantum gases \cite{Gericke2007,Wuertz2009}. Using all-optical techniques we produce a Bose-Einstein condensate of $10^5$ rubidium atoms in a single beam optical dipole trap. The 2D optical lattice is generated by two independent laser systems. For each axis, a laser beam is split into two beams which interfere under an angle of $90^\circ$ at the position of the dipole trap. In this geometry, the laser wavelength of $850\,$nm results in a lattice spacing of 600\,nm in both directions. The condensate is adiabatically transferred into the ground state of the combined optical potential ramping up the laser power of the lattice beams in $200\,$ms. We can achieve a maximum lattice depth of 18 recoil energies, and the filling factor in the central part of the gas is $80$ atoms per site. Each site has a tube-like shape with an extension of 6\,$\mu$m along the direction of the electron beam. During the imaging procedure the electron beam is scanned across the cloud. Electron impact ionization produces ions which are detected by a channeltron. The overall detection efficiency is about 15\,\%. Figure\,\ref{fig:figure1} shows a scanning electron microscope image of a Bose-Einstein condensate in the lattice potential. The image was obtained by summing over 260 individual experimental realizations.

\begin{figure}[th]
	\centering
		\includegraphics[width=0.47\textwidth]{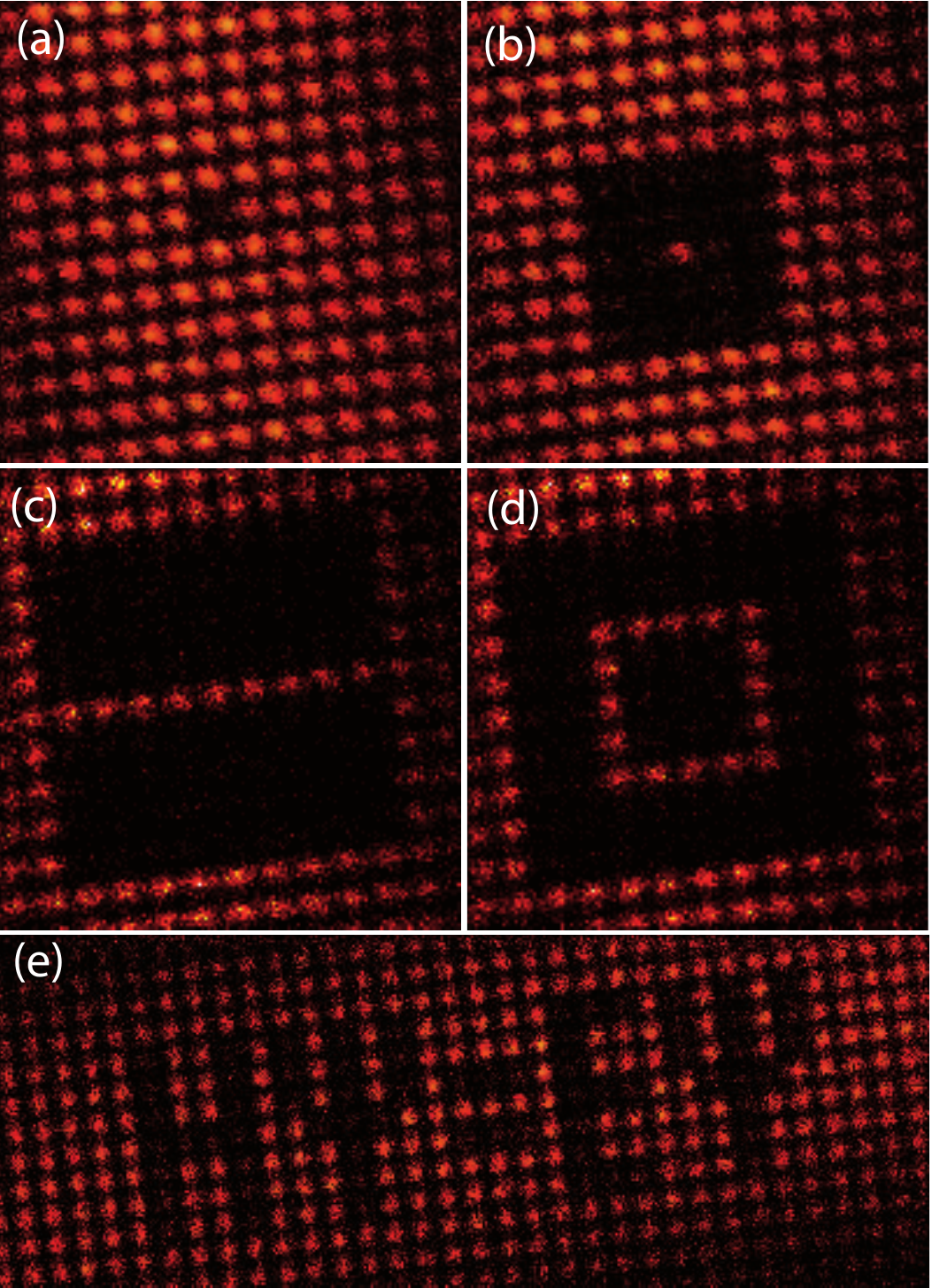}
	\caption{Patterning a Bose-Einstein condensate in a 2D optical lattice with a spacing of 600\,nm. Every emptied site was illuminated with the electron beam (7\,nA beam current, 100\,nm FWHM beam diameter) for 3\,ms (a,b),  2\,ms (c,d), and 1,5\,ms (e), respectively. The imaging time was 45\,ms. Between 150 and 250 images from individual experimental realizations have been summed for each pattern.}
	\label{fig:figure2}
\end{figure}

The addressing of individual sites in the optical lattice works as follows: the electron beam is pointed at selected sites for a dwell time of 1-3\,ms per site in order to remove the atoms. Immediately after the preparation, the imaging procedure is started. Four elementary examples of this patterning technique are presented in Fig.\,\ref{fig:figure2}. A single defect in the lattice structure is shown in Fig.\,\ref{fig:figure2}(a). The structure resembles a Schottky defect in a solid and is an ideal starting point to study the tunnelling dynamics close to a defect. The opposite situation corresponds to an isolated lattice site and is shown in Fig.\,\ref{fig:figure2}(b). Such a mesoscopic ensemble provides, for instance, the possibility to study the transition from few-body systems to the thermodynamic limit. It can also act as a paradigm for Rydberg blockade studies as the spatial extension of the ensemble is very small. In this context, quantum optical applications such as single atom and single photon sources or the creation of GHZ-like states have been proposed \cite{Lukin2001}. A chain and a ring of lattice sites are shown in Fig.\,\ref{fig:figure2}(c) and \ref{fig:figure2}(d) in order to illustrate the large variety of achievable geometries. Obviously, our approach allows for any arbitrary pattern that fits to the underlying quadratic lattice geometry (Fig.\,\ref{fig:figure2}(e)). 

Avoiding crosstalk to neighboring sites during the addressing procedure imposes stringent requirements on the spatial resolution. This is in contrast to a pure imaging process, for which it has been shown that neighboring sites can be distinguished even with a resolution limit well above the lattice spacing \cite{Karski2009}. Hence, imaging lattice sites and manipulating lattice sites set in general different requirements on the spatial resolution. As in our case the diameter of the electron beam (100\,nm FWHM) is much smaller than the lattice spacing (600\,nm), the sites can be addressed {\it without} affecting neighboring sites (see Fig.\,\ref{fig:figure2}(a)). The presented scheme is especially attractive in combination with single site occupancy. This can be achieved, for instance, via a 2D Mott insulator state \cite{Chin2009}. Subsequent patterning of the system can then provide the necessary spatial access to further manipulate and read-out the system by high-fidelity optical techniques.

The depth of the optical lattice in Fig.\,\ref{fig:figure2} is 18 recoil energies ($E_r=\pi^2\hbar^2 /(2ml^2)$, with $m$ being the rubidium mass and $l=600\,$nm). For these parameters the single particle tunneling time $t=h/(8J)$ amounts to 20\,ms \cite{tunnelingtime}. Here, $h$ is Plancks constant and $J$ ist the tunneling matrix element. As the preparation time (144\,ms for the pattern in Fig.\,\ref{fig:figure2}(c)) plus the subsequent imaging duration (45\,ms) is ten times larger than $t$, one might expect that a substantial fraction of the atoms rapidly refills the sites after the preparation and no structure is visible in the images. However, we only find a small population in the addressed lattice sites (12\,\% in average). A somewhat higher fraction of 25\,\% is found for the single defect in Fig.\,\ref{fig:figure2}(a) - in accordance with what one might expect as atoms from four neighboring sites can tunnel into the defect. Isolated sites are stable as well. The atom number in the single site shown in Fig.\,\ref{fig:figure2}(b) only decays to 55\,\% of its initial value and for the chain and the ring (Figs.\,\ref{fig:figure2}(c) and (d)) we find even higher values of 80\,\% - 90\,\% for the remaining population. These results suggest that tunneling in the patterned gas is highly supressed. 

In order to explain this observation we have to look at the system in more detail. The radial (axial) oscillation frequency in each tube is $\omega_r=2\pi\times12\,$kHz ($\omega_a=2\pi\times170\,$Hz). With a temperature of $T=100\,$nK and a chemical potential of $\mu=h\times2$\,kHz the gas in each tube is one-dimensional. While the atoms are radially in the ground state they occupy more than ten vibrational modes in the axial direction. The central line density amounts to $n_{\mathrm 1D}=15\,\mu m^{-1}$ and results in an interaction strength of $\gamma=0.08$. The dynamical evolution of the system after its preparation therefore corresponds to that of an array of coupled, weakly interacting 1D Bose gases. An atom tunneling from a filled site with $N$ atoms into an empty site has to overcome the total interaction energy $NU$, where $U$ is the on-site interaction energy. As $NU$ equals the chemical potential $\mu$, such a tunneling process is highly off-resonant and cannot be compensated by the small tunneling coupling $J$. Consequently, the tunneling dynamics is suppressed. While tunneling of all $N$ particles together is energetically allowed, it is highly unlikely as it requires a $N$th-order tunneling process. In fact, the tunneling dynamics in the patterned gas has similarities with nonlinear self-trapping \cite{Albiez2005} and the observation of repulsively bound pairs \cite{Winkler2006}. However, as the axial level splitting is much smaller than the interaction energy, tunneling into vibrationally excited axial modes of an empty tube might occur as an additional relaxation process. Moreover, as each tube can be described in local density approximation with an effective chemical potential $\mu_{\mathrm{eff}}(z)=\mu-\frac{1}{2}m\omega_a^2z^2$, where $z$ denotes the direction along the tube, the above reasoning fails close to the edges of each tube. These effects might explain why we observe a fraction of atoms refilling the empty tubes. We plan to extend these studies in the future by changing the lattice geometry and preparing the tubes perpendicular to the electron beam. This allows to additionally measure the axial extension in each tube and to analyze thermalization processes during the tunneling dynamics.

\begin{figure}
	\centering
		\includegraphics[width=0.47\textwidth]{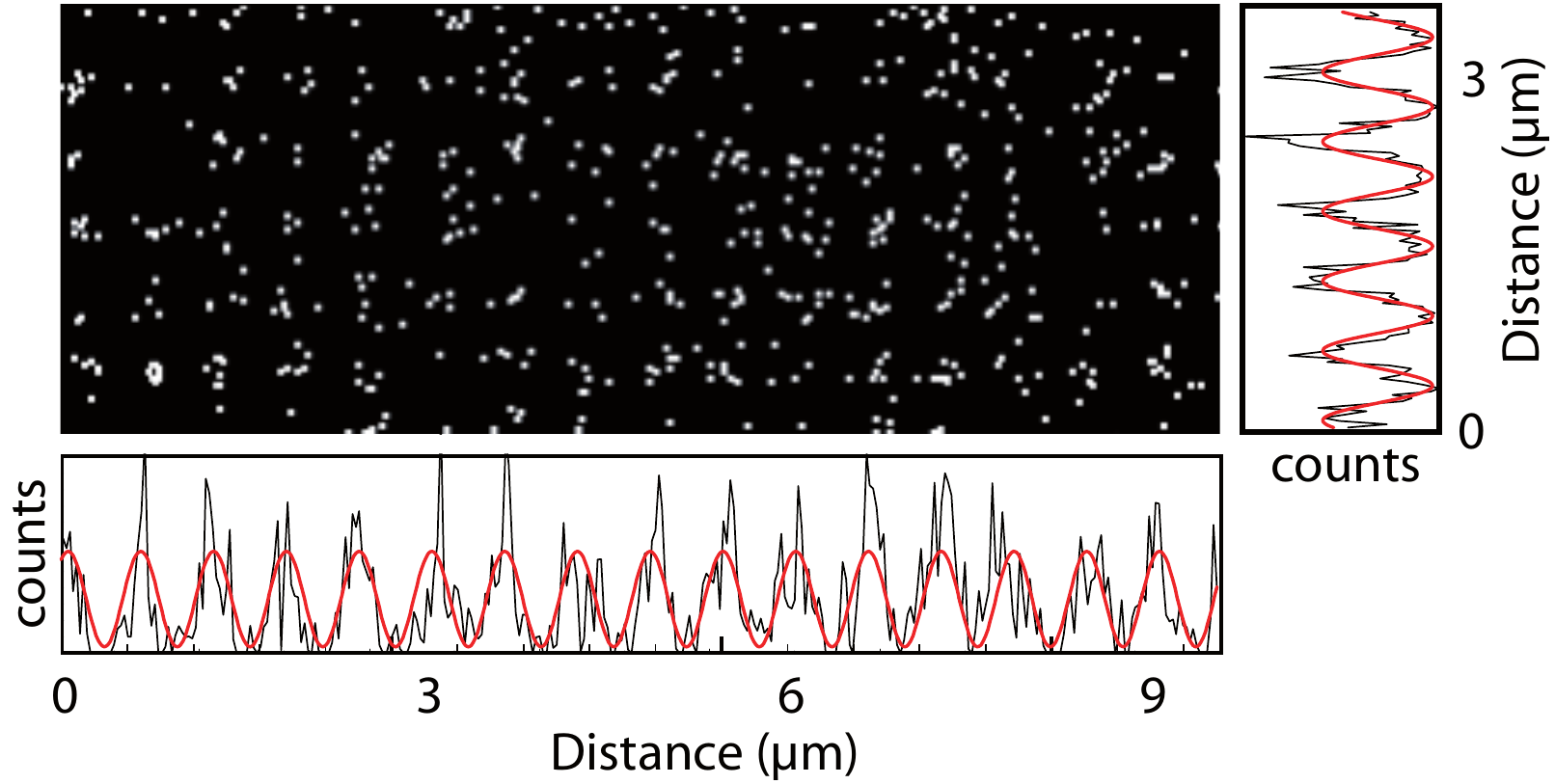}
	\caption{Single-shot image of the central part of the cloud. Each dot corresponds to a detected ion (800 in total). The integrated data along both lattice axes are shown on the side (black line). A fit to the data (red line) is used to determine the position of the lattice with an uncertainty of $\pm 30$\,nm.}
	\label{fig:figure3}
\end{figure}

\begin{figure}[t]
	\centering
		\includegraphics[width=0.450\textwidth]{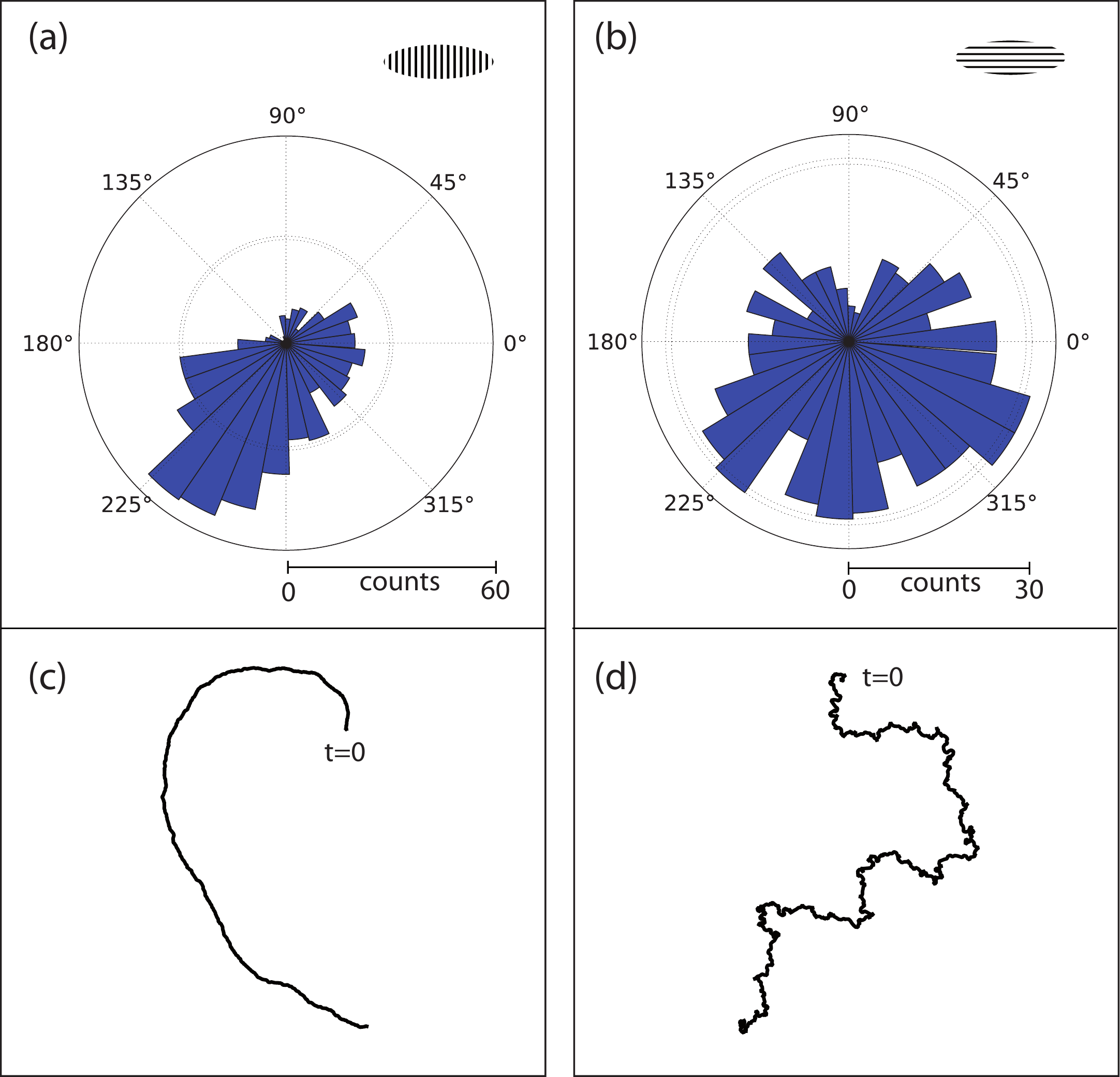}
	\caption{Drifts of the optical lattice. (a,b) Appearance of lattice positions (translated into a phase between $0$ and $2\pi$) for 300 experimental runs, corresponding to 1.5 hours of measurement time. The lattice vector is parallel (a) or perpendicular (b) to the condensate axis. (c,d) Illustration of the drift dynamics for the data in (a) and (b). Each lattice position is converted into a unity vector whose direction is given by the corresponding phase. All 300 vectors are successively plotted. A perfectly stable lattice would result in a straight line. Small long term drifts appear as large circles, while short term fluctuations lead to a pronounced zigzag shape. }
	\label{fig:figure4}
\end{figure}

The optical lattice can exhibit short and long term drifts in both axes. We account for this by phase analyzing each individual image. In Fig.\,\ref{fig:figure3} we show a single-shot image together with the integrated linescans in both lattice directions. From a fit to the linescan we extract the actual phase of the lattice and displace the image accordingly prior to the summation. We routinely correct all images applying this procedure. In order to illustrate the drift motion of the lattice we show in Fig.\,\ref{fig:figure4}(a) and (b) the phase distribution of a sequence of 300 images. While one axis is relatively stable over the whole sequence (corresponding to a total measurement time of 1.5 hours) the second axis shows an almost uniform distribution. This becomes even more evident if we plot the phase evolution of each lattice axis as a connected chain of unity vectors, where the direction of each unity vector represents the actual phase of the optical lattice (Fig.\,\ref{fig:figure4}(c) and (d)). We attribute the short term fluctuations of the lattice to mechanical vibrations of the setup and air turbulances in both arms of each lattice axis. Long term drifts are probably due to thermalization processes of the setup. We have checked that drifts of the field of view due to electrostatic charging in the electron column play a minor role. The observation of larger fluctuations and drifts for one axis is compliant with a less stable opto-mechanical layout of this axis. 

\begin{figure}[t]
	\centering
		\includegraphics[width=0.47\textwidth]{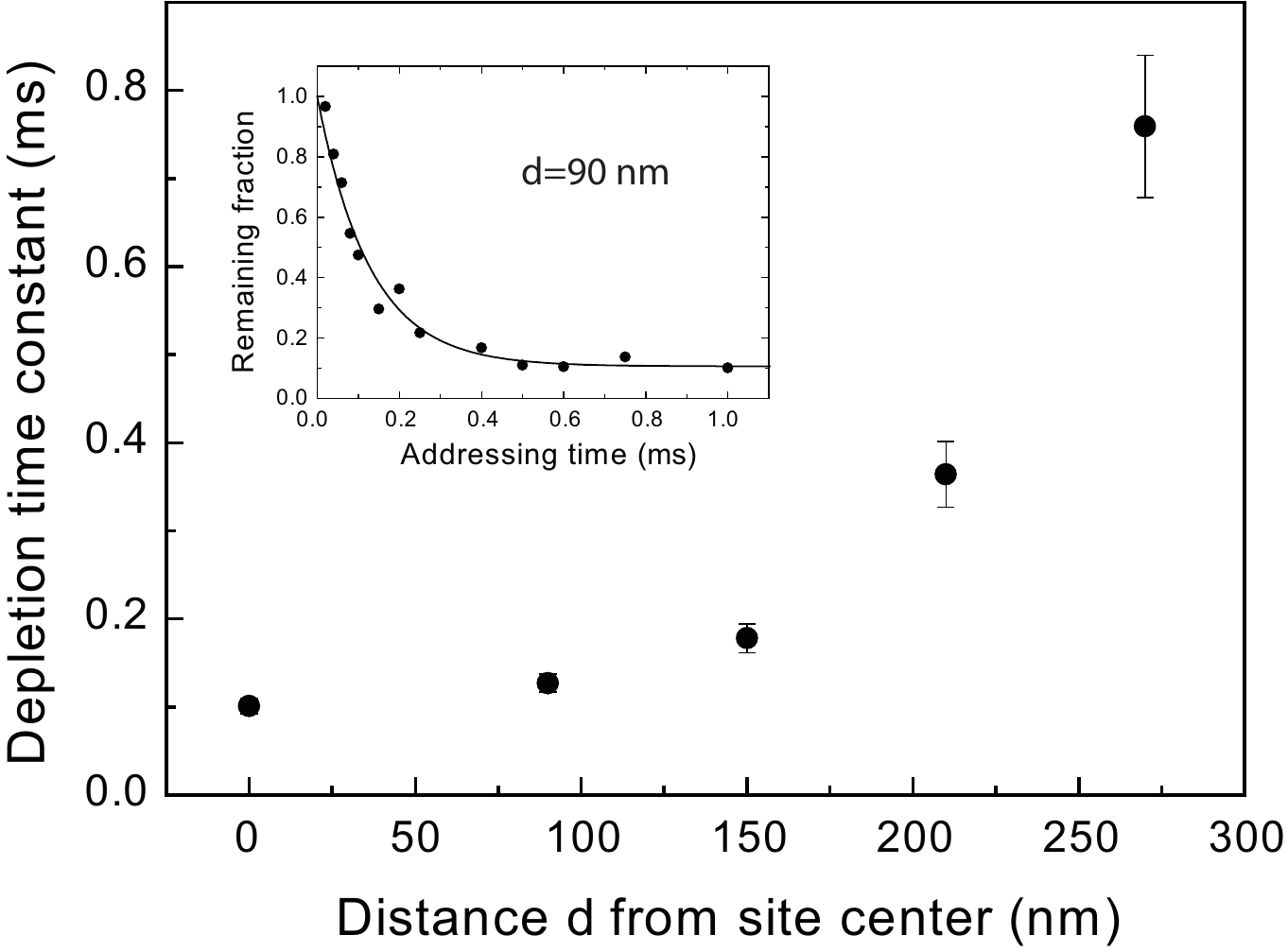}
	\caption{Depletion time constant for different positions of the electron beam with respect to the lattice site. At all distances we find an exponential decay. For central hits, the population decays with a time constant of 100 $\mu$s.  The inset shows the depletion at a distance of 90\,nm. The solid line is a fit with an exponential decay. The offset of 10\,\% represents atoms that are refilling the lattice site between the end of the preparation and the instant of imaging.}
	\label{fig:figure5}
\end{figure}

For the addressing scheme, the situation is more involved as a motion of the optical lattice can lead to an error in the addressing protocol which cannot be compensated for in a post processing procedure. This is the case, for instance, if the electron beam points in between two sites, thus affecting both of them. Due to the high spatial resolution the addressing procedure even works for some error in the lattice position. For the images in Fig.\,\ref{fig:figure2} we have accepted all phases in an interval $\pm\,0.6\times\pi$. A phase of $\pi$ corresponds to the extreme case where the electron beam points in between two sites. With this acceptance interval, we have to discard on average 65\,\% of the images. Imperfect addressing also leads to longer depletion times. The dependence of the depletion time constant on the position of the electron beam is shown in Fig.\,\ref{fig:figure5}. Whereas for central hits we find a time constant of $100\,\mu$s, it is more than 10 times larger if the electron beam points at the edge of a site. For all positions we find an exponential decay of the atom number. In terms of writing time, a lattice site can be prepared in about one millisecond - provided that the site is hit in the center \cite{preparation}. If we allow for a preparation time of 100\,ms a total of 100 sites can be emptied. This is sufficient to tailor a large part of the gas. 

The ability to remove atoms from specific sites has a direct possible application for the study of quantum phases in optical lattices. In many cases, the temperature of the gas after loading in the lattice potential is too high to reach the ground state of the system. Therefore, further cooling in the lattice is necessary. One possible cooling strategy relies on the spatial separation of the gas into regions of high and low entropy which are subsequently decoupled by a potential barrier from each other \cite{Bernier2009}. Our technique would directly allow to remove atoms from regions of high entropy without the need to decouple them from the remaining system.

In conclusion we have shown that arbitrary patterns of mesoscopic atomic ensembles in an optical lattice can be created and imaged by means of scanning electron microscopy. The patterns are remarkably stable even for a moderate lattice depth and allow for {\it in situ} studies of tunneling processes in optical lattices. The technique provides large flexibility and can be applied to any atomic species, mixture of different species or molecules. The tailored quantum gases can be further processed by standard manipulation and detection techniques and constitute a novel resource for cold atom physics. Ultimately, they might allow for the implementation of new schemes for quantum optical applications, quantum simulation and quantum information processing.

\begin{acknowledgments}
This work was supported by the DFG within the Emmy Noether-program and the University of Mainz.
\end{acknowledgments}


\begin{references}


\bibitem{Dalfovo1999}
F. Dalfovo, S. Giorgini, L. P. Pitaevskii, and S. Stringari, Rev. Mod. Phys. {\bf 71}, 463 (1999). 
\bibitem{Ketterle2008}
W. Ketterle and M. W. Zwierlein. In {\it Proceedings of the International School of Physics  Enrico Fermi}, eds. M. Inguscio, W. Ketterle, and C. Salomon (Amsterdam, IOS Press, 2008);
\bibitem{Ginsburg2007}
N. S. Ginsburg, S. R. Garner, and L. V. Hau, Nature {\bf 445}, 623 (2007).
\bibitem{Bloch2008}
I. Bloch, J. Dalibard, and W. Zwerger, Rev. Mod. Phys. {\bf 80}, 885 (2008).
\bibitem{Brennen1999}
G. K. Brennen, C. M. Caves, P. S. Jessen, and I. H. Deutsch, Phys. Rev. Lett. {\bf 82}, 1060 (1999).
\bibitem{Raussendorf2001}
R. Raussendorf, and H. J. Briegel, Phys. Rev. Lett. {\bf 86}, 5188 (2001).
\bibitem{Urban2008}
E. Urban {\it et al.}, Nature Physics {\bf 5}, 110 (2009).
\bibitem{Gaetan2008}
A. Gaetan {\it et al.}, Nature Physics {\bf 5}, 115 (2009).
\bibitem{Lengwenus2006}
A. Lengwenus, J. Kruse, M. Volk, W. Ertmer, and G. Birkl, Appl. Phys. B {\bf 86}, 377 (2006).
\bibitem{Karski2009} M. Karski {\it et al.}, Phys. Rev. Lett. {\bf 102}, 053001 (2009).
\bibitem{Fortagh2007}
J. Fort{\'a}gh and C. Zimmermann, Rev. Mod. Phys. {\bf 79}, 235 (2007).
\bibitem{Whitlock2009}
S. Whitlock, R. Gerritsma, T. Fernholz, and R. J. C. Spreeuw, New. J. Phys. {\bf 11}, 023021 (2009).
\bibitem{Greiner2002}
M. Greiner, O. Mandel. T. Esslinger, T. W. H{\"a}nsch, and I. Bloch, Nature {\bf 415}, 39 (2002).
\bibitem{Zhang2006}
C. Zhang, S. L. Rolston, and S. Das Sarma, Phys. Rev. A {\bf 74}, 042316 (2006).
\bibitem{Gorshkov2008}
A. V. Gorshkov, L. Jiang, M. Greiner, P. Zoller, and M. D. Lukin, Phys. Rev. Lett. {\bf 100}, 093005 (2008).
\bibitem{Kollath2007}
C. Kollath, M. K{\"o}hl, and T. Giamarchi, Phys. Rev. A {\bf 76}, 063602 (2007). 
\bibitem{Nelson2007}
K. Nelson, X. Li, and D. Weiss, Nature Physics {\bf 3}, 556 (2007).
\bibitem{Gericke2008}
T. Gericke, P. W{\"u}rtz, D. Reitz, T. Langen, and H. Ott, Nature Physics, {\bf 4}, 949 (2008).
\bibitem{Gericke2007}
T. Gericke, P. W{\"u}rtz, D. Reitz, C. Utfeld, and H. Ott, Appl. Phys. B {\bf 89}, 447 (2007)
\bibitem{Wuertz2009}
P. W{\"u}rtz, T. Gericke, T. Langen, A. Koglbauer and H. Ott, J. Phys.: Conf. Ser. {\bf 141}, 012020 (2008).
\bibitem{Lukin2001}
M. D. Lukin {\it et al.}, Phys. Rev. Lett. {\bf 87}, 037901 (2001).
\bibitem{Chin2009}
N. Gemelke, X. Zhang, C.-L. Hung, and C. Chin, arXiv:0904.1532.  
\bibitem{tunnelingtime}
A single atom released from a site in a 1D optical lattice periodically tunnels through the lattice with partial revivals at each lattice site separated by the characteristic tunneling time $h/4J$. Between two revivals in the same lattice site, the probability amplitude of the atom goes to zero and we identify $h/8J$ as the single particle tunneling time in a 1D lattice. In a 2D lattice, the revivals are more peaked but still occur with a periodicity of $h/4J$. We therefore take $h/8J$ as the reference tunneling time for our system.
\bibitem{Albiez2005}
M. Albiez {\it et al.}, Phys. Rev. Lett. {\bf 95}, 010402 (2005).
\bibitem{Winkler2006}
K. Winkler {\it et al.}, Nature {\bf 441}, 853 (2006).
\bibitem{preparation}
This is not a fundamental restriction. An improved electron column should be capable to achieve a 10 times higher beam current resulting in a 10 times faster writing time.
\bibitem{Bernier2009}
J. Bernier {\it et al.}, Phys. Rev. A {\bf 79}, 061601 (2009).
\end{references}
\end{document}